\documentclass[11pt]{myllncs}
\usepackage[T1]{fontenc}
\usepackage[latin1]{inputenc}
\usepackage[english]{babel}
\usepackage[fleqn]{amsmath}
\usepackage{a4wide,amssymb,epsfig,latexsym,multicol,array,hhline}
\usepackage{gastex,enumerate}

\def\N{\mathbb{N}}

\def\P0{{\rm P}_0}

\def\TERMS{\mathbb{T}}

\newcommand{\deducesem}{\:|\!\!\!=\!\!\!\!=\:}

\newenvironment{prooof}[0]{\noindent{\bf Proof}:}{$\Box$\\}

\newcommand{\ACT}{{\rm ACT}}
\newcommand{\INDSTR}{{\rm INDSTR}}
\newcommand{\LABA}{{\rm LAB}_{\cal A}}
\newcommand{\LABT}{{\rm LAB}_{\cal T}}
\newcommand{\NEXT}{{\rm NEXT}}
\newcommand{\PLAYS}{{\rm PLAYS}}
\newcommand{\PREF}{{\rm PREF}}
\newcommand{\SUCC}{{\rm SUCC}}
\newcommand{\reduito}[2]{\stackrel{#1}{\longrightarrow_{#2}}}
\begin{document}
\begin{center}
{\Large Jancar's formal system for deciding bisimulation of first-order grammars\\ 
and its non-soundness.}\\
by G\'eraud S\'enizergues\\
LaBRI and Universit\'e de Bordeaux I
\footnote{
mailing adress:LaBRI and UFR Math-info, Universit\'e Bordeaux1\\
               351 Cours de la lib\'eration -33405- Talence Cedex.\\
email:ges@labri.u-bordeaux.fr; 
fax: 05-40-00-66-69;\\
URL:http://dept-info.labri.u-bordeaux.fr/$\sim$ges/
}
%\thanks{This work has been partly supported by the BIG project;}
\end{center}

\paragraph{Abstract}:
We construct an example of proof within the main formal system from \cite{Jan10}, which
is intended to capture the bisimulation equivalence for non-deterministic first-order grammars,
and show that its conclusion is semantically false.
We then locate and analyze the flawed argument in the soundness (meta)-proof of \cite{Jan10}.\\
%\vfill
\pagenumbering{arabic}
\tableofcontents
\section{The grammar}
We consider the alphabet of actions ${\cal A}$, an intermediate alphabet of 
labels ${\cal T}$
and a map $\LABA: {\cal T} \rightarrow {\cal A}$ defined by:
$${\cal T} := \{ x,y,z,\ell_1\},\;\;{\cal A} := \{a,b,\ell_1\},\;\;\mbox{ and }$$
$$\LABA: x \mapsto a,\;\;y \mapsto a,\;\; z \mapsto b,\;\; \ell_1 \mapsto \ell_1.$$
(these intermediate objects ${\cal T}$, $\LABA$ will ease the definition of $\ACT$ below).
We define a first-order grammar ${\cal G} = ({\cal N},{\cal A},{\cal R})$
by: 
$${\cal N} := \{A, A', A'', B, B', B'', C, D, E, L_1\}$$
and the set of rules ${\cal R}$ consists of the following:
\begin{eqnarray}
A(v) &\reduito{y}{} & C(v)\\ %r1
A(v) &\reduito{x}{} & A'(v)\\%r2
B(v) &\reduito{x}{} & C(v)\\%r3
B(v) &\reduito{y}{} & B'(v)\\%r4
C(v) &\reduito{x}{} & D(v)\\%r5
C(v) &\reduito{y}{} & E(v)\\%r6
A'(v) &\reduito{x}{} & A''(v)\\%r7
B'(v) &\reduito{x}{} & B''(v)\\%r8
A''(v) &\reduito{x}{} & D(v)\\%r9
B''(v) &\reduito{x}{} & E(v)\\%r10
D(v) &\reduito{x}{} & v \label{ruleD}\\%r11
E(v) &\reduito{x}{} & v \label{ruleE1}\\%r12
E(v) &\reduito{z}{} & v \label{ruleE2}\\%r13
L_1  &\reduito{\ell_1}{} & \bot \label{ruleL1}%r14
%D'(v) &\reduito{x}{} & v\\%r15
%E'(v) &\reduito{x}{} & v\\%r16
%E'(v) &\reduito{z}{} & v %r17
\end{eqnarray}
Let us name rule $r_i$ (for $1 \leq i \leq 14$), the rule appearing in order $i$ 
in the above list. 
We define a map $\LABT: {\cal R} \rightarrow {\cal T}$ by: $\LABT(r_i)$ is the terminal letter
used 
by the given rule $r_i$.
Subsequently we define $\ACT(r_i):= \LABA(\LABT(r_i))$.
Namely, $\ACT$ maps all the rules $r_1, \ldots , r_{12}$ onto $a$, $r_{13}$ on $b$
and $r_{14}$ on $\ell_1$.
\section{The formal system}
We consider the formal systems ${\cal J}(T_0,T'_0,S_0,{\cal B})$ defined in page 22 of \cite{Jan10}, which are intended to be sound and complete for the bisimulation-problem
for non-deterministic first-order grammars.
Let us denote by $\TERMS$ the set of all terms over the ranked alphabet 
${\cal N} \cup \{L_i\mid i \in \N\} \cup\{\bot\}$ (here the symbols $L_i$ have arity $0$).
\subsection{Prefixes of strategies}
The notion of {\em finite prefix of a D-strategy} is mentionned p. 23, line 11.
We assume it has the following meaning
\begin{definition}
Let $T,T' \in \TERMS$.
A  finite prefix of a D-strategy w.r.t. $(T,T')$ is a 
subset $S \subseteq ({\cal R}\times{\cal R})^*$ 
of the form
$$S = S'\cap ({\cal R}\times{\cal R})^{\leq n}$$
for some $n \in \N$ and some D-strategy $S'$ w.r.t. $(T,T')$.
\label{def-PDstrategy}
\end{definition}
In order to make clear that the above notion is effective, we consider the following notion of
D-q-strategy (Defender's quasi-strategy). 
\begin{definition}
Let $T,T' \in \TERMS$.
A {\em D-q-strategy} w.r.t. $(T,T')$ is a subset $S \subseteq ({\cal R}\times{\cal R})^*$ 
such that:\\
DQ1: $(\varepsilon,\varepsilon) \in S$\\
DQ2: $S$ is prefix-closed\\
DQ3: $S\subseteq \PLAYS(T,T')$\\
DQ4: $\forall \alpha \in S$,\\
either $\alpha \backslash S=\{(\varepsilon,\varepsilon)\}$\\ 
or $\NEXT((T,T'),\alpha) \notin \sim_1$ \\
or [$\NEXT((T,T'),\alpha) \in \sim_1$ and the set $\{ (\pi,\pi') \in {\cal R}\times{\cal R} \mid \alpha\cdot (\pi,\pi')\in S\}$ is full for $\NEXT((T,T'),\alpha)$].
\label{def-Dqstrategy}
\end{definition}
Note that a D{\em-strategy} is a D-q-strategy where, condition DQ4 is replaced by:\\
DQ'4: $\forall \alpha \in S$,\\
$\NEXT((T,T'),\alpha) \notin \sim_1$ \\
or [$\NEXT((T,T'),\alpha) \in \sim_1$ and the set $\{ (\pi,\pi') \in {\cal R}\times{\cal R} \mid \alpha\cdot (\pi,\pi')\in S\}$ is full for $\NEXT((T,T'),\alpha)$].\\
A {\em winning} D-strategy, is a D-q-strategy where condition DQ4 is replaced by:\\
DQ''4: $\forall \alpha \in S$,\\
$\NEXT((T,T'),\alpha) \in \sim_1$ and the set $\{ (\pi,\pi') \in {\cal R}\times{\cal R} \mid \alpha\cdot (\pi,\pi')\in S\}$ is full for $\NEXT((T,T'),\alpha)$.\\
\begin{lemma}
Every finite prefix of a strategy is a D-q-strategy.
\label{L-PD_implies_DQ}
\end{lemma}
\begin{prooof}
Let $S'$ be a D-strategy w.r.t. $(T,T')$ and
$$S= S'\cap ({\cal R}\times{\cal R})^{\leq n}$$
for some $n \in \N$, $S'$ D-strategy w.r.t. $(T,T')$.\\
DQ1: Since $S'$ is non-empty and prefix-closed $(\varepsilon,\varepsilon) \in S'$,
hence $(\varepsilon,\varepsilon) \in S'\cap S({\cal R}\times{\cal R})^{\leq n}$.\\
DQ2: $S'$ and  $({\cal R}\times{\cal R})^{\leq n}$ are both prefix-closed, hence their intersection
is also prefix-closed.\\
DQ3: $S'\subseteq \PLAYS(T,T')$ and $S \subseteq S'$, hence $S\subseteq \PLAYS(T,T')$\\
DQ4: $\forall \alpha \in S$,\\
$\NEXT((T,T'),\alpha) \notin \sim_1$ \\
or [$\NEXT((T,T'),\alpha) \in \sim_1$ and the set $\{ (\pi,\pi') \in {\cal R}\times{\cal R} \mid \alpha\cdot (\pi,\pi')\in S'\}$ is full for $\NEXT((T,T'),\alpha)$].
If $|\alpha| < n$, the above property holds in $S$.\\
If $|\alpha| = n$, the property $\alpha \backslash S=\{(\varepsilon,\varepsilon)\}$ holds.
In all cases DQ4 is fulfilled.\\
\end{prooof}
\begin{definition}
We define the {\em extension} ordering over ${\cal P}(({\cal R}\times{\cal R})^*)$ as follows:
for every $S_1,S_2 \in {\cal P}(({\cal R}\times{\cal R})^*)$, $S_1 \sqsubseteq S_2$ iff the two conditions below hold:\\
E1- $S_1 \subseteq S_2$\\
E2- $\forall \alpha \in S_2-S_1, \exists \beta \in S_1, \mbox{ which is maximal in  } S_1 
\mbox{ for the prefix ordering and such that },\\
\beta \preceq \alpha.$
\label{def-extension}
\end{definition}
\begin{lemma}
Let $T,T'\in \TERMS$.
The extension ordering over the set  of all D-q-strategies w.r.t. $(T,T')$, is inductive.
\label{L_inclusion_is_inductive}
\end{lemma}
\begin{prooof}
We recall that a partial  order $\leq$ over a  set $E$ is {\em inductive} iff, 
every totally ordered subset of $E$ has some upper-bound.\\
One can check that, if $P$ is a set of D-q-strategies w.r.t. $(T,T')$, which is totally ordered by $\sqsubseteq$, then the set
$$ S := \bigcup_{s \in P} s $$
is still a D-q-strategy and fulfills:
$$ \forall s \in P, s \sqsubseteq S.$$
Hence the extension ordering over the set of D-q-strategies w.r.t. $(T,T')$ is inductive. 
\end{prooof}
\begin{lemma}
Let $S \subseteq ({\cal R}\times{\cal R})^*$ be finite and let $n:= \max\{ |\alpha| \mid
\alpha \in S\}$.\\
$S$ is a finite prefix of a 
D-strategy w.r.t. $(T,T')$ iff\\
(1) $S$ is a D-q-strategy w.r.t. $(T,T')$\\
(2) $\forall \beta \in S, [ \beta \backslash S = \{(\varepsilon,\varepsilon)\} 
\Rightarrow (|\beta| = n \mbox{ or } \NEXT((T,T'),\beta) \notin \sim_1] $).
\label{L-characterisation_PDstrategies}
\end{lemma}
\begin{prooof}
{\bf Direct implication}:\\
Let $S'$ be a D-strategy w.r.t. $(T,T')$ and
$$S= S'\cap ({\cal R}\times{\cal R})^{\leq n}$$
for some $n \in \N$ and some  $S'$ which is a D-strategy w.r.t. $(T,T')$.\\
1- By Lemma \ref{L-PD_implies_DQ} $S$ is a D-q-strategy w.r.t. $(T,T')$.\\
2- Suppose that $\beta \in S, \beta \backslash S = \{(\varepsilon,\varepsilon)\}$ and 
$|\beta| < n$. Then $\beta \backslash S' = \{(\varepsilon,\varepsilon)\}$ too.
Since $S'$ is a D-strategy w.r.t. $(T,T')$, this implies that 
$\NEXT((T,T').\beta) \notin \sim_1$.\\
{\bf Converse}:\\
Suppose that $S$ fulfills conditions (1)(2).
By Lemma \ref{L_inclusion_is_inductive},  Zorn's lemma applies on the set of 
D-q-strategies w.r.t. $(T,T')$: there exists a maximal D-q-strategy $S'$ 
(for the extension ordering) such that
$S \sqsubseteq S'$.
Since $S'$ is maximal, if $\alpha \in S'$ and $\alpha \backslash S=\{(\varepsilon,\varepsilon)\}$,
$\NEXT((T,T'),\alpha) \notin \sim_1$.
Thus, instead of the weak property DQ4, $S'$ fulfills the property:
$$\forall \alpha \in S',
\NEXT((T,T'),\alpha) \notin \sim_1\;\;
\mbox{ or }$$ 
$$[\NEXT((T,T'),\alpha) \in \sim_1 \mbox{ and }\{ (\pi,\pi') \in {\cal R}\times{\cal R} \mid \alpha\cdot (\pi,\pi')\in S\} \mbox{ is full for } \NEXT((T,T'),\alpha)].$$
Hence $S'$ is a strategy w.r.t. $(T,T')$.\\
Clearly
$$ S \subseteq S' \cap ({\cal R}\times{\cal R})^{\leq n}.$$
Let us prove the reverse inclusion.\\
Let $\alpha \in S' \cap ({\cal R}\times{\cal R})^{\leq n}$.
Let $\beta$ be the longuest word in $\PREF(\alpha) \cap S$.\\
If $\beta = \alpha$, then $\alpha \in S$, as required.\\
Otherwise $\alpha \in S'-S$. By condition E2 of definition \ref{def-extension}, there exists some 
$\beta \in S$, which is maximal in $S$ for the prefix ordering and such that
$$\beta \prec \alpha.$$
Maximality of $\beta$ implies, by condition (2) of the lemma, that, 
$$|\beta| = n \mbox{ or } \NEXT((T,T').\beta) \notin \sim_1.$$
Since $\beta \prec \alpha$ we are sure that $|\beta| < n$ so that 
$$\NEXT((T,T').\beta) \notin \sim_1.$$
This last statement contradicts the fact that $\beta \backslash S'$ is a D-strategy, w.r.t
$\NEXT((T,T').\beta)$ which is 
non-reduced to $\{(\varepsilon,\varepsilon)\}$ (since it posesses $\beta^{-1}\alpha$).\\
We can conclude that $\alpha \in S$.
Finally:
$$S = S' \cap ({\cal R}\times{\cal R})^{\leq n}.$$
\end{prooof}
\begin{lemma} Let $T,T'\in \TERMS$ and let $S \subseteq ({\cal R}\times{\cal R})^*$ be finite.
One can check whether  
$S$ is a finite prefix of a 
D-strategy w.r.t. $(T,T')$ 
\label{L-decidability_PDstrategies}
\end{lemma}
This follows immediately from the characterisation given by Lemma \ref{L-characterisation_PDstrategies}.
\subsection{Formal systems}
\label{subsec_formal_systems}
For every $T_0,T'_0 \in \TERMS$, $S_0$ finite prefix of strategy w.r.t $(T_0,T_0)$
 and finite ${\cal B} \subseteq \TERMS \times \TERMS,$ is defined a formal system
$${\cal J}(T_0,T'_0,S_0,{\cal B})$$
The set of judgments of all the systems are the same.
But the axiom and one rule (namely R7), is depending on the parameters $(T_0,T'_0,S_0,{\cal B})$.
\subsection{Judgments}
A {\em judgment} has one of the three forms:\\
{\bf FORM 1}:\\
$$m \deducesem (T,T',S)$$
where $m \in \N$, and $T,T'\in \TERMS$ are regular terms and $S$ is a finite prefix of a strategy.
w.r.t. $(T,T')$ (D-strategies are defined p.20, lines 27-30;
finite prefixes are mentionned, though in a fuzzy way. at p. 23, line 11; we shall apply 
here Definition \ref{def-PDstrategy}).\\
{\bf FORM 2}:\\
$$ m \deducesem (T,T',S) \leadsto \alpha \deducesem (T_1,T'_1,S_1)$$ where $m \in \N$, $(T,T',S), (T_1,T'_1,S_1)$ fulfilling the above conditions,
$\alpha \in S$ and $\alpha \backslash S = S_1$.\\
{\bf FORM 3}:\\
$$ m \deducesem (T,T',S) \leadsto \alpha \deducesem SUCC$$
where  $m \in \N$, $(T,T',S)$ fulfill  the above conditions and $\alpha \in S$.\\
For all systems ${\cal J}(T_0,T'_0,S_0,{\cal B})$ the set of judgments is the same and consists of
all the items of one of the three above forms.
\subsection{Basis}
We call {\em basis} every finite set
$${\cal B} \subseteq \TERMS \times \TERMS.$$
\subsection{Axioms}
${\cal J}(T_0,T'_0,S_0,{\cal B})$ has a single axiom:
$$0 \deducesem (T_0,T'_0,S_0)$$
\subsection{Deduction rules}
All the systems ${\cal J}(T_0,T'_0,S_0,{\cal B})$
have the set of rules described page 22 of \cite{Jan10}. 
We name them $R1,R2, \ldots, R10$, the number
corresponding to the one in the text. Note that R7 depends on the basis ${\cal B}$.
\subsection{Proofs}
\noindent Let $T_0,T'_0 \in \TERMS$.
A {\em proof} of $T_0 \sim T'_0$ within the family of formal systems defined above is
a finite basis ${\cal B}$, together with, for each $(T,T') \in {\cal B} \cup \{(T_0,T'_0)\}$
a finite prefix of D-strategy $S$ w.r.t. $(T,T')$ and a proof, within system ${\cal J}(T,T',S,{\cal B})$ of the judgment 
$$ 0 \deducesem (T,T',S) \leadsto (\varepsilon,\varepsilon) \deducesem {\rm SUCC}.$$
\section{The Equivalence proof}
\label{sec-equivalence_proof}
We exhibit here a proof of 
$$A(\bot) \sim B(\bot).$$
According to the above notion of proof, it consists of the following items.\\
\noindent Basis:
$${\cal B} := \{(C(L_1),C(L_1)), (D(L_1),D(L_1)), (E(L_1),E(L_1))\}.$$
Proofs:\\
\begin{itemize}
\item a proof of the judgment $0 \deducesem A(\bot),B(\bot),S\leadsto (\varepsilon,\varepsilon) \deducesem {\rm SUCC}$ in the formal system ${\cal J}(A(\bot),B(\bot),S,{\cal B})$ (see $\pi_3$).
\item a proof of the judgment $0 \deducesem C(L_1),C(L_1),{\rm Id}_{C,1}\leadsto 
(\varepsilon,\varepsilon) \deducesem {\rm SUCC}$ in the formal system ${\cal J}(C(L_1),C(L_1),{\rm Id}_{C,1},{\cal B})$ (see $\pi_4$).
\item a proof of the judgment $0 \deducesem D(L_1),D(L_1),{\rm Id}_{D,2}\leadsto (\varepsilon,\varepsilon) \deducesem {\rm SUCC}$ in the formal system ${\cal J}(D(L_1),D(L_1),{\rm Id}_{D,2},{\cal B})$ (see $\pi_5$).
\item a proof of the judgment $0 \deducesem E(L_1),E(L_1),{\rm Id}_{E,2}\leadsto (\varepsilon,\varepsilon) \deducesem {\rm SUCC}$ in the formal system ${\cal J}(E(L_1),E(L_1),{\rm Id}_{D,2},{\cal B})$ (see $\pi_6$). 
\end{itemize}
\begin{figure}[htbf]
      \begin{center} 
        \includegraphics[width=14cm]{example.41.eps}
      \end{center}
\caption{The proof $\pi_1$}
\end{figure}
where $H(A,B)$ stands for $0 \deducesem A(\bot),B(\bot), S$.\\
Proof $\pi_2$:
\begin{figure}[htbf]
      \begin{center} 
        \includegraphics[width=10cm]{example.42.eps}
      \end{center}
\caption{The proof $\pi_2$}
    \end{figure}
\begin{figure}[htbf]
      \begin{center} 
        \includegraphics[width=10cm]{example.43.eps}
\caption{The proof $\pi_3$}
      \end{center}
    \end{figure}
\begin{figure}[htbf]
      \begin{center} 
        \includegraphics[width=16cm]{example.44.eps}
      \end{center}
\caption{The proof $\pi_4$}
    \end{figure}
where $H(C,C)$ stands for $0 \deducesem C(L1),C(L1),{\rm Id}_{C,1}$.\\
\begin{figure}[htbf]
      \begin{center} 
        \includegraphics[width=10cm]{example.45.eps}
      \end{center}
\caption{The proof $\pi_5$}
    \end{figure}
\begin{figure}[htbf]
      \begin{center} 
        \includegraphics[width=14cm]{example.46.eps}
      \end{center}
\caption{The proof $\pi_6$}
    \end{figure}
In the above proofs the following defender strategies 
(or prefix of strategies) were used
(in fact, they can be deduced from the proofs):\\
Let 
$${\cal S} := \{(yx,xy),(yy,xx),(xxx,yxx)\}.$$
For every subset $Z$ of $({\cal A} \times {\cal A})^* $, by $\PREF(Z)$
we denote its set of prefixes.\\
We define
$${\cal P} := \PREF({\cal S})$$
namely:
$${\cal P} = \{(\varepsilon,\varepsilon),(y,x),(yx,xy),(x,y),(xx,yx),(xxx,yxx)\}$$
Finally, we define $S$ as the subset of $({\cal R} \times {\cal R})^* $ obtained by 
replacing, in ${\cal P}$, every 2-tuple $(u,v) \in ({\cal A} \times {\cal A})^* $ by the
unique 2-tuple  $(r_u,r_v) \in ({\cal R} \times {\cal R})^*$, such that
$r_u$ (resp. $r_v$) is applicable on $A$ (resp. on $B$), $\LABT(r_u) = u$ and $\LABT(r_v) = v$.
Namely:
$$S = \{ (\varepsilon,\varepsilon),(r_1,r_2),(r_1r_5,r_2r_6),(r_1r_6,r_2r_5),(r_2,r_1), (r_2r_7,r_1r_8),(r_2r_7r_9,r_1r_8r_{10})\}.$$
(See figures \ref{figure_stratST}-\ref{figure_stratSR}).\\
Subsequently:
\begin{eqnarray*}
S_1 & := &\{(\varepsilon,\varepsilon),(r_5,r_6),(r_6,r_5)\}\\
S_2 & := &\{(\varepsilon,\varepsilon)\}\\
S_3 & := &\{ (\varepsilon,\varepsilon),(r_7,r_8),(r_7r_9,r_8r_{10})\}\\
S_4 & := &\{ (\varepsilon,\varepsilon),(r_9,r_{10})\}\\
S_5 & := &\{ (\varepsilon,\varepsilon)\}\\
S_6 & := & \INDSTR(S_2,S_5)= S_2^{-1} \circ S_5= \{ (\varepsilon,\varepsilon)\}
\end{eqnarray*}
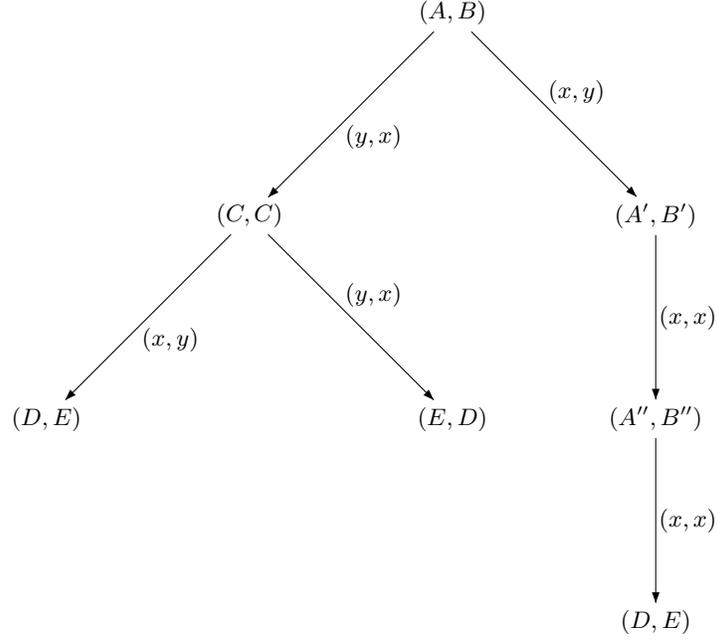
\begin{figure}[t]
\begin{center}
\setlength{\unitlength}{.9mm}
\begin{picture}(90,90)
\gasset{Nframe=n,Nadjust=wh,Nadjustdist=1,AHnb=1,ELdist=0.5}
\node(DE)(0,30){$(D,E)$}
\node(ED)(60,30){$(E,D)$}
\node(CC)(30,60){$(C,C)$}
\node(AB)(60,90){$(A,B)$}
\node(APBP)(90,60){$(A',B')$}
\node(ASBS)(90,30){$(A'',B'')$}
\node(DEBOT)(90,0){$(D,E)$}
\drawedge(AB,CC){$(y,x)$}
\drawedge(CC,DE){$(x,y)$}
\drawedge(CC,ED){$(y,x)$}
\drawedge(AB,APBP){$(x,y)$}
\drawedge(APBP,ASBS){$(x,x)$}
\drawedge(ASBS,DEBOT){$(x,x)$}
\end{picture}
\end{center}
\caption{The strategy $S$ viewed on ${\cal T}$}
\label{figure_stratST}        
\end{figure}

\begin{figure}[t]
\begin{center}
\setlength{\unitlength}{.9mm}
\begin{picture}(90,90)
\gasset{Nframe=n,Nadjust=wh,Nadjustdist=1,AHnb=1,ELdist=0.5}
\node(DE)(0,30){$(D,E)$}
\node(ED)(60,30){$(E,D)$}
\node(CC)(30,60){$(C,C)$}
\node(AB)(60,90){$(A,B)$}
\node(APBP)(90,60){$(A',B')$}
\node(ASBS)(90,30){$(A'',B'')$}
\node(DEBOT)(90,0){$(D,E)$}
\drawedge(AB,CC){$(r_1,r_3)$}
\drawedge(CC,DE){$(r_5,r_6)$}
\drawedge(CC,ED){$(r_6,r_5)$}
\drawedge(AB,APBP){$(r_2,r_4)$}
\drawedge(APBP,ASBS){$(r_7,r_8)$}
\drawedge(ASBS,DEBOT){$(r_9,r_{10})$}
\end{picture}
\end{center}
\caption{The strategy $S$}
\label{figure_stratSR}        
\end{figure}
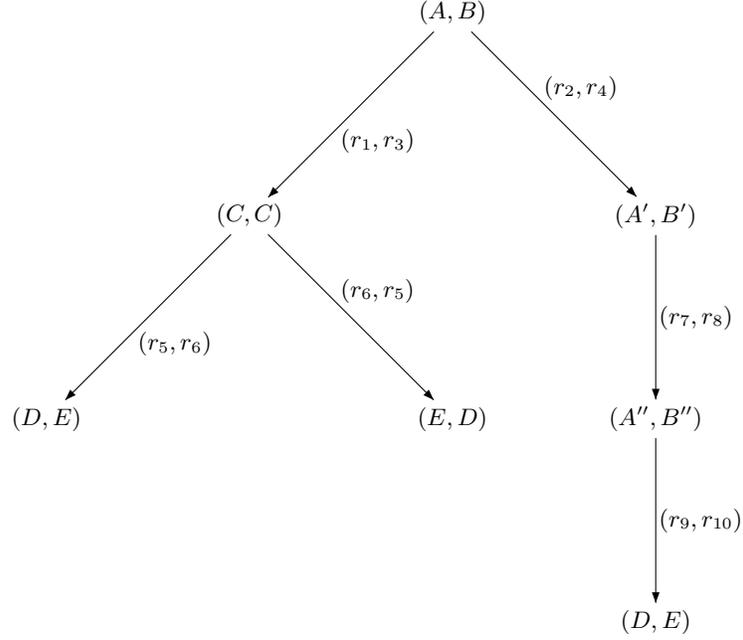
\begin{lemma}
$S$ is a prefix of D-strategy w.r.t. $(A(\bot),B(\bot))$.
\label{S_is_strategy}
\end{lemma}
\begin{prooof}
Let us check that $S$ fulfills the critetium given by 
Lemma \ref{L-characterisation_PDstrategies}. Here $n=3$.
Point (1) is easily checked.\\
Let $\beta \in ({\cal R}\times{\cal R})^{*}$ such that $\beta \backslash S = \{(\varepsilon,\varepsilon)\}$. Either ($\NEXT((A,B),\beta)\in \{(E,D),(D,E)\}$, while $D \not{\!\!\sim_1} E$) 
or $|\beta| = 3$.
Hence Point (2) holds.
\end{prooof}
For proving the equivalences of the members of the basis we shall use
the ``trivial'' prefixes of strategies, consisting of 2-tuples of identical rules on both sides:
\begin{eqnarray*}
{\rm Id_{C,1}} & := & \{(\varepsilon,\varepsilon), (r_5,r_5),(r_6,r_6)\})\\
{\rm Id_{D,2}} & := & \{(\varepsilon,\varepsilon), (r_{11},r_{11}), (r_{11}r_{14},r_{11}r_{14})\})\\
{\rm Id_{E,2}} & := & \{(\varepsilon,\varepsilon), (r_{12},r_{12}), (r_{13},r_{13}),(r_{12}r_{14},r_{12}r_{14}), (r_{13}r_{14},r_{13}r_{14})\}).
\end{eqnarray*}
One can check that ${\rm Id_{C,1}}$ is a prefix of the strategy, for the game with initial 
position $(C,C)$,
$${\rm Id}_{C,\infty}:=\{(u,u) \mid u \in {\cal R}^*, C(L_1) \reduito{u}{}\}.$$
The set ${\rm Id_{D,2}}$ (resp. ${\rm Id_{E,2}}$) is really a strategy for the game with 
initial position $(D,D)$ (resp. $(E,E)$) since no rule $r_i$ is applicable on $\bot$.
For every $N \in \{C,D,E\}$, the symbol ${\rm Id_{N,i}}$ will denote a residual of 
length $i$ of the strategy ${\rm Id_{N,n}}$:
\begin{eqnarray*}
{\rm Id_{C,0}} & = & {\rm Id_{D,0}} = {\rm Id_{E,0}} = \{(\varepsilon,\varepsilon)\},\\
{\rm Id_{D,1}} & = & {\rm Id_{E,1}} = \{(\varepsilon,\varepsilon), (r_{14},r_{14})\}
\end{eqnarray*}
\section{The Non-equivalence (meta-) proof}
\begin{lemma}
$A(\bot) \not{\!\!\sim} B(\bot)$
\label{lem-nonequivalence_proof}
\end{lemma}
\begin{prooof}
$$\forall u \in {\cal R}^*, ACT(u) = aaab \Rightarrow A(\bot) \not{\!\!\reduito{u}{}}$$
while
$$\exists u \in {\cal R}^*, ACT(u) = aaab \mbox { and } B(\bot) \reduito{u}{}$$
hence $A(\bot) \not{\!\!\sim} B(\bot)$.
\end{prooof}
From section \ref{sec-equivalence_proof} and Lemma \ref{lem-nonequivalence_proof} we conclude
\begin{theorem}
The family of formal systems $({\cal J}(T_0,T'_0,S_0,{\cal B}))$ is not sound.
\end{theorem}
\section{Variations}
Let us describe variations around this example.\\
\paragraph{Description of the proofs}$\;\;$\\
We chosed to write the proofs with judgments of the form $m \deducesem (T,T',S)$ or $m \deducesem (T,T',S) \leadsto \alpha \deducesem (T_1,T'_1,S_1)$ or $m \deducesem (T,T',S) \leadsto \alpha \deducesem \SUCC$, where, in the case of forms 2,3, the prefix $\alpha$ is given by its image under 
the map $\LABT$ (its image is enough to determine $\alpha \in ({\cal R} \times {\cal R})^*$
just because the grammar is deterministic).
Of course the proofs can be rewritten with prefixes $\alpha \in ({\cal R} \times {\cal R})^*$.
\paragraph{Strategies}$\;\;$\\
The formal systems ${\cal J}(T_0,T'_0,S_0,{\cal B})$ described in subsection \ref{subsec_formal_systems} were devised so that their set of judgments is recursive.
Let us consider now the formal systems $\hat{{\cal J}}(T_0,T'_0,S_0,{\cal B})$ really considered 
in pages 21-24. Their judgments are also of the forms 
$$m \deducesem (T,T',S),\;\;m \deducesem (T,T',S) \leadsto \alpha \deducesem (T_1,T'_1,S_1),\;\; m \deducesem (T,T',S) \leadsto \alpha \deducesem \SUCC$$
but where $S,S_1$ are D-strategies (instead of finite prefixes of strategies),
``except when a judgment is obtained by rule R2'': see the fuzzy remark on page 23, line 11,
followed by the enigmatic remark that ``we could complete the definition anyhow for such cases''.
Since $S,S_1,S_2,S_3,S_4,S_5,{\rm Id_{D,2}},{\rm Id_{E,2}}$ are really D-strategies and $S_6$ is obtained by an application of
rule R2, it seems that our proofs $\pi_3,\pi_5,\pi_6$ are also proofs in the systems $\hat{{\cal J}}(T_0,T'_0,S_0,{\cal B})$. As well, replacing ${\rm Id_{C,1}}$ by ${\rm Id_{C,\infty}}$ 
in $\pi_4$, we obtain a proof of judgment 
$0 \deducesem (C(L_1),C(L_1),{\rm Id_{C,\infty}}) \leadsto (\varepsilon,\varepsilon) \deducesem \SUCC$ in the system $(\hat{{\cal J}}(C(L_1),C(L_1),{\rm Id_{C,\infty}},{\cal B}))$.
\paragraph{Depth of the examples}$\;\;$\\
One can devise such proofs of non-bisimilar pairs, with an arbitrary long initial strategy:
it suffices to add non-terminals $D_1,D_2,\ldots, D_k,E_1,E_2,\ldots ,E_k$ and to replace 
rules (\ref{ruleD},\ref{ruleE1},\ref{ruleE2},\ref{ruleL1}) by the sequence of rules:
\begin{eqnarray}
D(v) &\reduito{x}{} & D_1(v) \label{nruleD}\\%r11
E(v) &\reduito{x}{} & E_1(v) \label{nruleE}\\%r12
\vdots&&\vdots \nonumber\\
D_1(v) &\reduito{x}{} & D_2(v) \label{nruleD1}\\%r13
E_1(v) &\reduito{x}{} & E_2(v) \label{nruleE1}\\%r14
\vdots&&\vdots \nonumber\\
D_k(v) &\reduito{x}{} & v \label{nruleDk}\\%r
E_k(v) &\reduito{x}{} & v \label{nruleEk}\\%r
E_k(v) &\reduito{z}{} & v \label{nruleEkz}\\%r
L_1  &\reduito{\ell_1}{} & \bot %
\end{eqnarray}
A proof of $0 \deducesem A(\bot),B(\bot),\hat{S} \leadsto (\varepsilon,\varepsilon) \deducesem {\rm SUCC}$ can still be written, but with a longer initial strategy 
$\hat{S}$ where the maximal length of words is $3+k$, and a prefix of strategy $\hat{S}_6$ 
of length $k$. Note that the sizes of the proofs $\pi_3,\pi_4,\pi_5,\pi_6$ still remain the same.
\section{The flawed argument}
Let us locate precisely, in \cite{Jan10}, the crucial {\em flawed} argument in 
favor of soundness of the systems.\\
Page 24, line \$-4, the following assertion (FA) is written:\\
``The final rule in deriving
$m \deducesem (U,U',S') \leadsto (\varepsilon,\varepsilon) \deducesem SUCC$ could not be the Basis rule, due to the least eq-level assumption for $T,T'$ (recall Prop. 17)''.\\

In our example:
$$(T,T') = (A(\bot),B(\bot)),\;\;
EqLv((A(\bot),B(\bot))=3$$
Let us take
$$(U,U',S')= (E(\bot),E(\bot),S_6)$$
We have:
$$EqLv(U,U',S') =0 = EqLv(T,T',S)-3$$
And the judgment 
$$3 \deducesem E(\bot),E(\bot),S_6 \leadsto (\varepsilon,\varepsilon) \deducesem SUCC$$
can be derived by the proof $\pi_7$ below.
\begin{figure}[htbf]
      \begin{center} 
        \includegraphics[width=14cm]{example.47.eps}
\caption{The proof $\pi_7$}
      \end{center}
    \end{figure}
Hence $(T,T')$ has the {\em least} equivalence level, among the EqLevels of 
the elements of $\{ (T,T') \} \cup {\cal B}$ while  $m,U,U'$ fulfills the {\em maximality} 
hypothesis of the text (line \$-7).\\
But the final rule used in this proof is the basis rule (R7), contradicting the assertion (FA).\\

The bug seems to be the following:
by Proposition 17
\begin{equation}
EqLv(E(L_1),E(L_1)) \leq EqLv(E(\bot),E(\bot))
\label{prop17}
\end{equation}
BUT
\begin{equation}
EqLv(E(L_1),E(L_1)) > EqLv(E(\bot),E(\bot),S_6)\;\; !
\label{true_inequality}
\end{equation}
A superficial look at the instance (\ref{prop17}) of Proposition 17 
can induce the idea that, for {\em every} D-strategy ${\cal S}$ (in particular for $S_6$), 
the inequality
\begin{equation}
EqLv(E(L_1),E(L_1)) \leq EqLv(E(\bot),E(\bot),{\cal S})
\label{false_inequation}
\end{equation} 
holds. 
In fact, what shows Proposition 17, is that inequality (\ref{false_inequation}) does hold but,
only for strategies ${\cal S}$ which are {\em optimal} for the defender, hence realizing exactly
the equivalence level of $(E(\bot),E(\bot))$.
%\bibliographystyle{alpha}
%\bibliography{example}

\end{document}